\documentclass[%
reprint,
notitlepage,
superscriptaddress,
tightenlines,
amsmath,amssymb,aps,
]{revtex4-1}

\usepackage{soul}
\usepackage{graphicx}
\usepackage{textcomp}
\usepackage{amsmath}
\usepackage{mathtools}
\usepackage{color}
\usepackage{xspace}
\usepackage{amsfonts}
\usepackage{mathrsfs}
\usepackage{gensymb}
\usepackage[normalem]{ulem}

\definecolor{darkblue}{rgb}{0,0,0.6}
\definecolor{darkred}{rgb}{0.6,0,0}
\definecolor{orange}{rgb}{0.9,0.5,0.25}
\usepackage[%
  colorlinks=true,
  urlcolor=darkblue,
  linkcolor=darkred,
  citecolor=darkblue
]{hyperref}

\draft 

\newcommand*\uDD{u_{\text{DD}}}
\newcommand*\uDT{u_{\text{DT}}}
\newcommand*\uE{u_{\text{E}}}
\newcommand*\uEr{u_{\text{Er}}}
\newcommand*\uEl{u_{\text{El}}}
\newcommand*\ue{u_{\text{e}}}
\newcommand*\ud{u_{\text{d}}}
\newcommand*\ude{u_{\text{de}}}

\newcommand*\deltabu{\mathbf{\delta u}}
\newcommand*\bu{\mathbf{u}}

\newcommand*\nD{n_{\text{D}}}
\newcommand*\nE{n_{\text{E}}}

\newcommand*\Dc{D_{\text{c}}}
\newcommand*\Dd{D_{\text{d}}}
\newcommand*\De{D_{\text{e}}}
\newcommand*\Dde{D_{\text{de}}}

\newcommand*\lamDD{\lambda_{\text{DD}}}
\newcommand*\kD{k_{\text{D}}}
\newcommand*\kdD{k_{\text{dD}}}
\newcommand*\kdE{k_{\text{dE}}}
\newcommand*\kdEr{k_{\text{dEr}}}
\newcommand*\kdEl{k_{\text{dEl}}}
\newcommand*\ked{k_{\text{ed}}}
\newcommand*\kde{k_{\text{de}}}
\newcommand*\ke{k_{\text{e}}}
\newcommand*\kd{k_{\text{d}}}

\newcommand*\bD{\mathbf{D}}

\newcommand*\bn{\mathbf{n}}

\begin{document}

\newcommand{\fig}{Fig. }
\newcommand{\eqn}{Eqn. }
\newcommand{\eqns}{Eqns. }
\newcommand{\rev}[1]{{\textbf{\color{black} #1} }}


\title{3D pattern formation of a protein-membrane suspension}

\author{Am\'elie Chardac}
\affiliation{Department of Physics, Brandeis University, 415 South street, Waltham, MA 02453}
\author{Michael M. Norton}
\affiliation{Department of Physics, Brandeis University, 415 South street, Waltham, MA 02453}
\author{Jonathan Touboul}
\affiliation{Department of Mathematics and Volen National Center for Complex Systems, Brandeis University, 415 South street, Waltham, MA 02453}
\author{Guillaume Duclos}
 \homepage{gduclos@brandeis.edu}
\affiliation{Department of Physics, Brandeis University, 415 South street, Waltham, MA 02453}

\begin{abstract}

Many essential cellular processes, including cell division and the establishment of cell polarity during embryogenesis, are regulated by pattern-forming proteins. These proteins often need to bind to a substrate, such as the cell membrane, onto which they interact and form two-dimensional (2D) patterns. It is unclear how the membrane's continuity and dimensionality impact pattern formation. Here, we address this gap using the MinDE system, a prototypical example of pattern-forming membrane proteins. We show that when the lipid substrate is fragmented into submicrometer-sized diffusive liposomes, ATP-driven protein-protein interactions generate three-dimensional (3D) spatially extended patterns, despite the complete loss of membrane continuity. Remarkably, these 3D patterns emerge at scales four orders of magnitude larger than the individual liposomes. By systematically varying protein concentration, liposome size, and density, we observed and characterized a variety of 3D dynamical patterns not seen on continuous 2D membranes, including traveling waves, dynamical spirals, and a coexistence phase. Simulations and linear stability analysis of a coarse-grained model revealed that the physical properties of the dispersed membrane effectively rescale both the protein-membrane binding rates and diffusion, two key parameters governing pattern formation and wavelength selection. These findings highlight the robustness of Min's pattern-forming ability, suggesting that protein-membrane suspensions could serve as an adaptable template for studying out-of-equilibrium self-organization in 3D, beyond \emph{in vivo} contexts. 

\end{abstract}

\maketitle

\section{Significance}

The spontaneous formation of patterns by multi-protein systems is essential for orchestrating fundamental biological processes like cell division and embryogenesis. Although various membrane proteins can generate patterns on continuous substrates, the mechanisms governing their self-organization on discontinuous, dispersed, and motile membranes, typical of intracellular environments, remain elusive. Here, we demonstrate that MinDE proteins, a prototypical pattern-forming system from \textit{E. coli}, can form robust 3D patterns when interacting with a suspension of nanoscale liposomes, rather than a continuous 2D lipid membrane. Through a combination of modeling and \textit{in vitro} experiments, we reveal that membrane fragmentation controls pattern formation by effectively rescaling biochemical and transport rates. Our results provide critical insights into intracellular self-organization and the design of tunable, out-of-equilibrium biomimetic systems.

\section{Introduction}

Pattern formation via reaction-diffusion processes is essential for biological self-organization across scales ~\cite{turing1990chemical,koch1994biological,kondo2010reaction,halatekSelforganizationPrinciplesIntracellular2018}. Intracellular protein patterns coordinate critical biological processes by coupling chemically fueled protein-protein interactions to passive or active transport processes. Importantly, the protein-protein interactions that mediate pattern formation are often hosted on a substrate, such as lipid membranes~\cite{Bement2015NCB,TurbulenceDefects_Fakhri2020,wigbers2021hierarchy, chen2024energy,Raskin1999, HU20011337,Szeto2003,Huang2003,Loose2008,etemad1995asymmetrically,ringgaard2009movement,Goehring2011,ietswaart2014competing} or DNA~\cite{ebersbach2008self,havey2012atp,Wang2014,merino2021self}. Considering the availability, geometry and mobility of the substrate is essential to understand how these pattern-forming systems self-organize. Although reaction-diffusion models of membrane proteins often assume continuity of a two-dimensional substrate, intracellular membranes are often fragmented, heterogeneous, and motile, raising the question of whether large-scale patterns can emerge in such environments. To address this question, we leverage the reconstituted MinDE protein system, replacing its usual planar lipid membrane with a suspension of submicrometer-spaced liposomes.\\

The spatiotemporal oscillations of the MinDE proteins on the membrane of \textit{E. Coli} has served as a paradigmatic model system to study pattern formation \textit{in vivo}~\cite{Raskin1999, Coyle2024, Szeto2003,HU20011337} and \textit{in vitro}~\cite{Loose2008, Vecchiarelli2014}. 
\textit{In vivo}, the pole-to-pole oscillations of MinD lead to the robust assembly of a cytokinetic ring of FtsZ polymers in the middle of the bacterium, ensuring the formation of two daughter cells of the same size~\cite{HU20011337}. \textit{In vitro}, MinDE complexes form a variety of dynamic and static patterns on continuous lipid membranes~\cite{Glock2019},  which can template the self-organization of other membrane-bound proteins~\cite{Ramm2018}, cargos~\cite{Ramm2021,Gavrilovic2024}, or drive the transport of cell-sized liposomes~\cite{fu2023}. 

\begin{figure*}
    \includegraphics[width=0.9\linewidth]{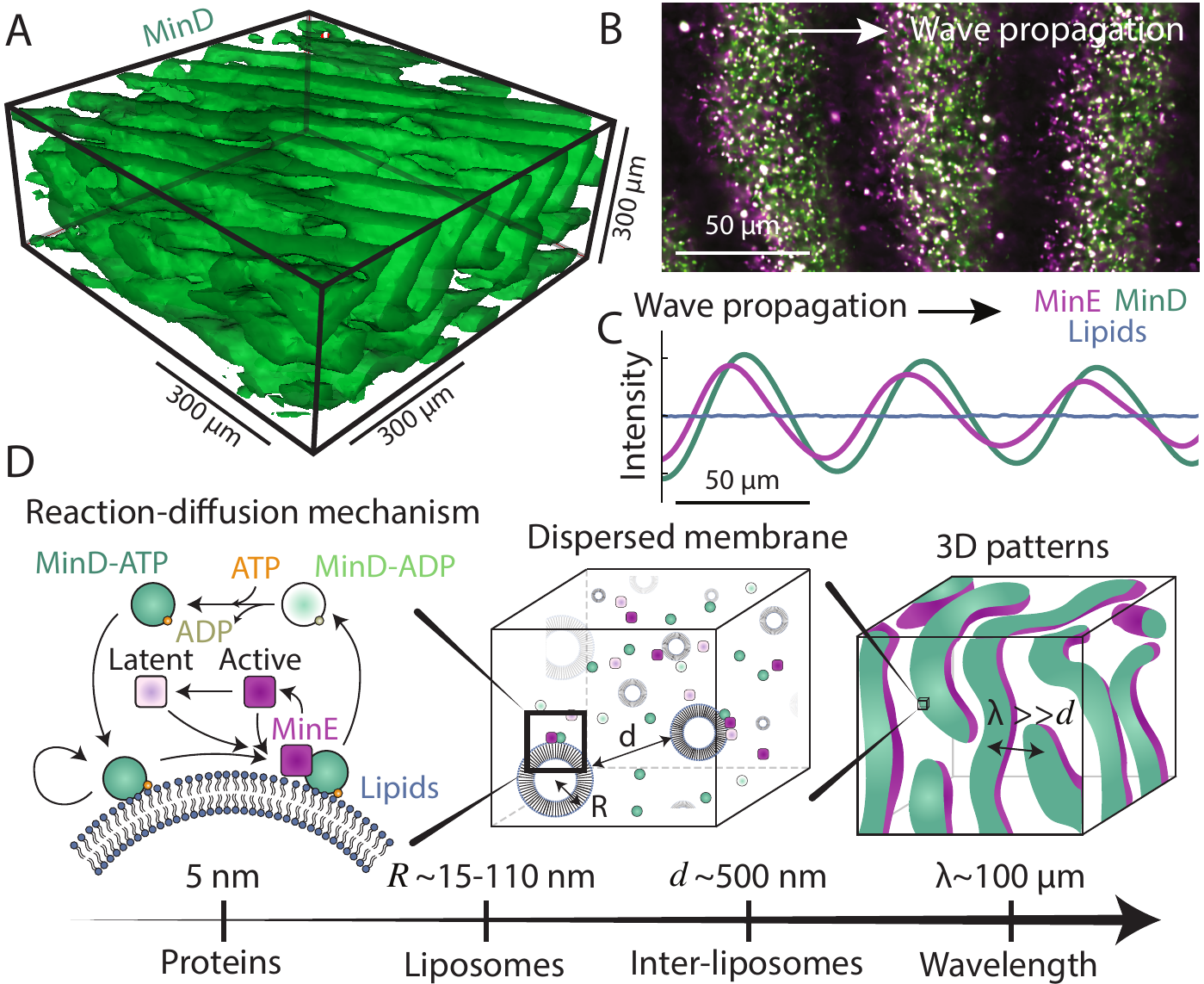}
    \caption{\textbf{The formation of 3D patterns on a dispersed membrane is a multiscale phenomena}. \textbf{(A)} 3D reconstruction of dynamical patterns of MinD proteins from confocal microscopy.  \textbf{(B)} Confocal z-slice of MinD (green) and MinE (magenta) forming traveling waves on the dispersed membrane. The largest lipid vesicles onto which proteins are bound are visible using fluorescence microscopy. The mean liposomes' radius is $\rm 15\,nm$. \textbf{(C)} Intensity profile of MinD (green), MinE (magenta), and the lipids (blue) along the direction of propagation of the wave (black arrow). MinE is located at the rear of the waves, following MinD. The lipids are spatially homogeneous. \textbf{(D)} Schematic of the chemical reaction network between MinD, MinE, ATP, and the dispersed membrane. The typical wavelength $\lambda$ of the traveling waves is significantly larger than the size R of individual liposomes and the inter-liposome distance $d$.}
    \label{fig:overview}
\end{figure*}

The biochemical mechanisms underlying pattern formation of Min proteins on a continuous two-dimensional lipid bilayer are well understood~\cite{halatekSelforganizationPrinciplesIntracellular2018, Loose2008, MinEswitch_Frey2018}. Briefly, upon adenosine triphosphate (ATP) binding, MinD spontaneously dimerizes and inserts itself into a lipid membrane. MinE binds to membrane-bound MinD, triggering ATP hydrolysis, and MinD detachment. MinE subsequently detaches and either rapidly binds to another membrane-bound MinD or transitions to a latent conformation with lower MinD affinity~\cite{MinEswitch_Frey2018}. The cycle is complete when MinD exchanges its bound-adenosine diphosphate (ADP) for another ATP (\fig\ref{fig:overview}D). In this classical picture, protein interactions with the membrane are crucial, and while pattern formation is certainly mediated by the cytosol surrounding the membrane ~\cite{ GeoSense_Schweizer2012,halatek2018,halatekSelforganizationPrinciplesIntracellular2018,brauns2021, MembraneBulkCoupling_Frey2021,brauns2021a,Wurthner2022,Gesele2020}, little remains known about the impact of the membrane's continuity on pattern formation.

In this paper, we leverage the cytosolic coupling to extend the Min system beyond its native configuration, examining its ability to form patterns in a 3D reactive fluid, where the membrane is not only discontinuous but dispersed and freely diffusing within a 3D fluid. Combining experiments, simulations, and linear stability analysis of a coarse-grained model, we reveal that MinDE proteins in a suspension of submicrometer-spaced liposomes form 3D patterns that are qualitatively distinct from the 2D patterns found on flat continuous membranes. Recasting the fragmented membrane as another reactive species in the Min system reveals how the physical properties of the substrate rescale both the reactive and diffusive rates of the membrane-bound Min proteins. Our findings expand the known strategies for controlling biological pattern formation, further establishing the MinDE oscillations as a paradigmatic \textit{in vitro} model system to study self-organization in 3D, despite only exhibiting 2D patterns \textit{in vivo}. Finally, our work demonstrate how to leverage protein-protein interactions outside of their native biological context to build biomolecular materials that bridge the gap between matter and life.

\section{3D large-scale patterns form on a suspension of submicrometer-sized liposomes.}\label{section:patterns}
To investigate the mechanisms governing pattern formation on a dispersed substrate, we used the \textit{in vitro} Min system from \textit{E. Coli}~\cite{Rowlett2013, denk2018a,Raskin1999}, which we modified to replace the continuous cell membrane with a homogeneous suspension of liposomes with a mean radius ranging from  {$15$ to $110\, \rm nm$ and a $15 \%$ polydispersity} (see Methods~\ref{section:MethodsExp}, \fig S1). Consequently, and in sharp contrast to \emph{in vivo} situations or \emph{in vitro} reconstitution of the system using continuous lipid bilayers, the substrate on which the protein-protein interaction occurs is discontinuous, curved, and dispersed in a three-dimensional volume where each liposome is free to diffuse isotropically.

\begin{figure*}[t]
    \includegraphics[width=1\linewidth]{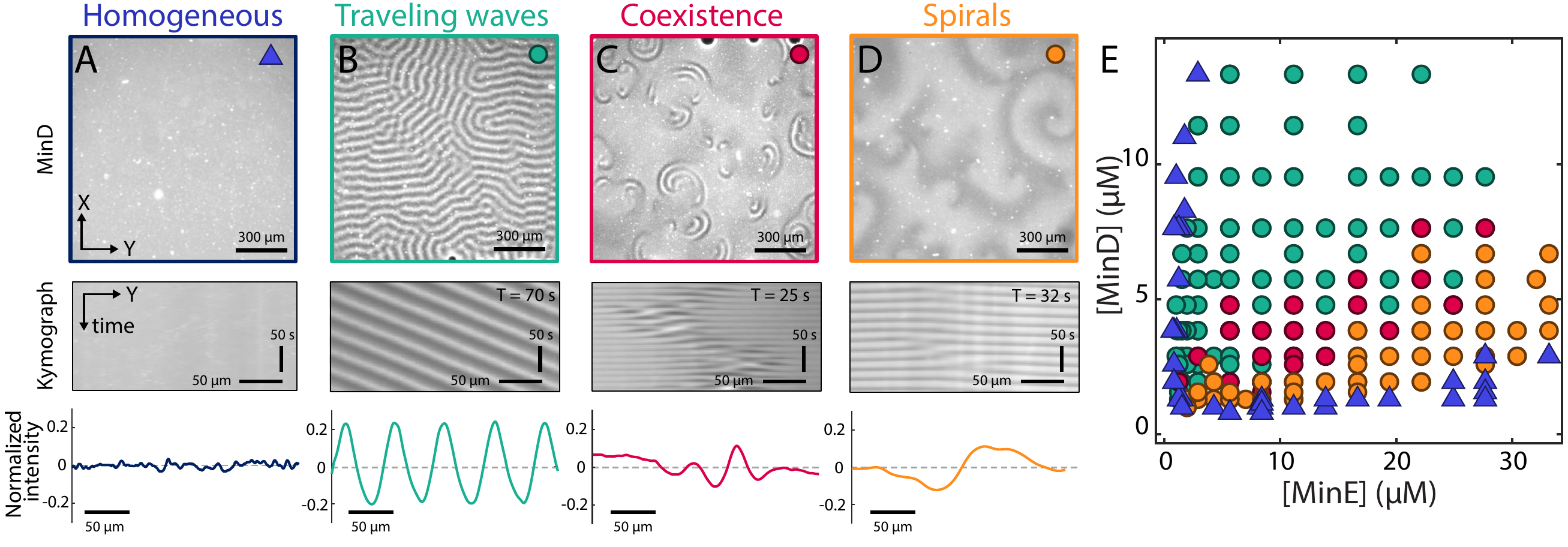}
    \caption{\label{fig:wide}\textbf{Proteins concentrations regulate pattern selection.} \textbf{(A)} Homogeneous steady-state, \textbf{(B)} Traveling waves, \textbf{(C)} Coexistence phase, \textbf{(D)} Spiral phase. The top row shows middle plane z-slices of MinD fluorescence intensity (Widefield microscopy), the middle row shows kymographs (space-time plots) of the pattern, and the bottom row shows the normalized MinD fluorescence intensity along a line. \textbf{(E)} Experimental phase diagram where the concentrations of MinD and MinE are changed. Each data point represents two to five replicates. For all those experiments, the liposomes suspension properties are kept constant: mean radius $R = 15\,\rm nm$ and concentration $ c = 0.4\, \rm mM$.}
    \label{fig:patterns}
\end{figure*}

Here, we report that an initially homogeneous mixture of MinE and MinD forms 3D dynamical patterns when interacting with a suspension of liposomes. Confocal microscopy revealed that MinD protein patterns form and propagate in all directions, spanning the system size (\fig\ref{fig:overview}A, Movie S1, \fig S2 and \fig S3). Fluorescence imaging of both proteins and lipids revealed two key observations: i) MinE trails behind MinD (\fig\ref{fig:overview}B-C), consistent with patterns on continuous membranes~\cite{Loose2011} and ii) the lipids remain homogeneously distributed in 3D (\fig\ref{fig:overview}C, \fig S4). Importantly, the length scales at play in this oscillatory reaction cover multiple orders of magnitude (\fig\ref{fig:overview}D): the wavelength of the patterns typically spans hundreds to thousands of vesicles. The ability to form patterns with a wavelength larger than the spacing between reaction sites prompted us to further identify the physical mechanisms and relevant lengthscales that contribute to the formation of dynamical patterns in 3D.

\begin{figure*}[t]
\includegraphics[width=0.84\linewidth]{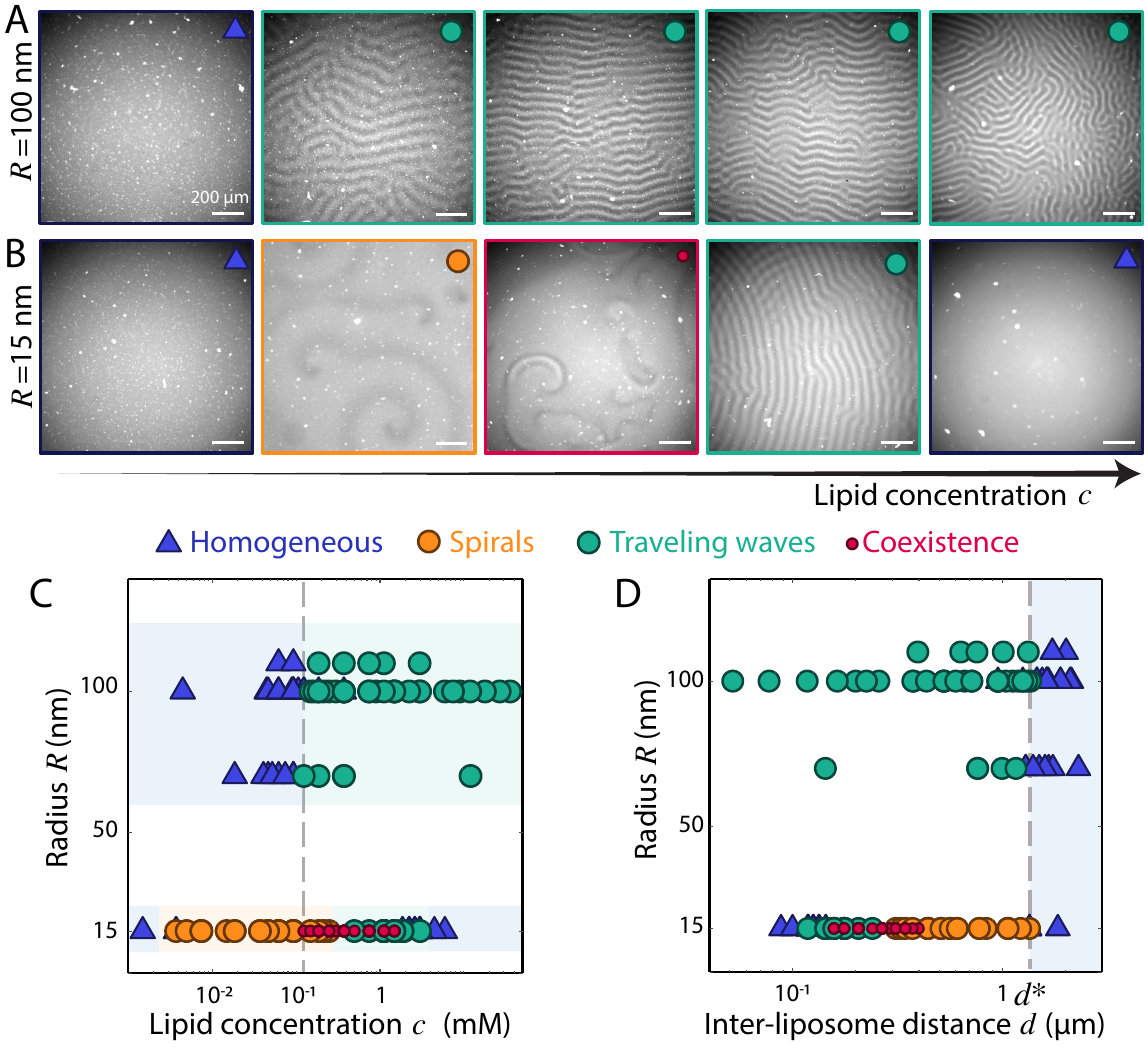}
    \caption{\textbf{The properties of the liposomes suspension regulate pattern formation.} \textbf{(A)-(B)} Experimental MinD fluorescence images for two liposomes radius (\textbf{(A)}: $R = 100\,\rm nm$, \textbf{(B)}: $R = 15\,\rm nm$), at increasing lipid concentration $c$ (from left to right). Scale bar: $\rm 200\, \mu m$.  \textbf{(C)} Phase diagram in ($R$, $c$), where $R$ is the mean liposome's radius and $c$ the lipid concentration. Colored backgrounds represent regions of predominance of each type of patterns.  \textbf{(D)} Phase diagram in ($R$, $d$) where $d$ is the typical distance between two liposomes. The blue region indicates the region where no pattern form. The dashed line indicates the critical inter-liposome distance $d^{*}$ above which no patterns form. For \textbf{(C)} and \textbf{(D)}, each experiment was replicated two to five times. For all the experiments reported in this figure, proteins concentrations were fixed to $\mathrm{[MinD]} = 5.7\, \mu \mathrm{M}$ and $\mathrm{[MinE]} = 8.25\, \mu \mathrm{M}$.}
    \label{fig:membraneExp}
\end{figure*}

\section{Proteins concentrations regulate pattern selection}
To explore the variety of patterns formed and their dependence of protein concentration, we classified the three types of patterns observed - traveling waves, spirals, and a coexistence phase - through their kymographs and intensity profiles (\fig\ref{fig:patterns}A-D). Titrating MinD and MinE concentrations, we observed that the chemical composition of the system regulated the existence and the nature of the patterns formed (\fig\ref{fig:patterns}E and \fig S8). First, we observed that a minimal concentration of both proteins was required to destabilize the homogeneous steady state (\fig\ref{fig:patterns}A,E and Movie  {S7}), which is consistent with previous studies of Min patterns on continuous lipids membranes~\cite{Loose2008,Glock2019}. Above this critical concentration, pattern formation is robust for a range of protein concentrations spanning over a decade. 

The relative concentrations of MinD and MinE also tightly control the type of pattern observed. In particular, we found that traveling waves dominated when MinD was in excess compared to MinE. In this phase, multiple domains of planar waves propagating in different directions coexist (\fig\ref{fig:patterns}B and Movie S2). The direction of propagation was discontinuous at the boundary between two domains (\fig S5 and Movie S3). These lines of singularities evolved over timescales much longer than the typical oscillation period (Movie S3).

When, instead, MinE was in excess compared to MinD, we found that spirals patterns were dominating  (\fig\ref{fig:patterns}D, Movie S4). The spiral waves  propagate from well-defined motile cores that could annihilate when the cores of two counter-rotating spirals met. Symmetrically, we found that new pairs of counter-rotating spirals could emerge and nucleate from a spiral-free background (\fig S6). These spiral waves are reminiscent of scroll waves in excitable media \cite{KEENER1988269}.

At intermediate MinD/E concentrations, a third phase emerged, characterized by a coexistence of the two dynamical patterns described above. It consists
of spatially restricted high-amplitude short-wavelength traveling waves propagating from the core of low-amplitude large-wavelength spiral waves (\fig\ref{fig:patterns}C and Movie S5). The spatially restricted planar wave patterns are highly dynamical and short-lived. Movies and kymograph analysis suggest that the spatially restricted traveling wave pattern forms because of an instability near the core of the spiral wave where the MinD density is maximum, that, momentarily and locally, brings the system into the traveling-wave regime of the ([MinD],[MinE]) phase space (\fig\ref{fig:patterns}E). Diffusion eventually smooths the amplitude of the wave, effectively confining the traveling waves close to the core of the spirals. Statistical analysis of the MinD fluorescence intensity in the coexistence phase confirmed the distinct nature of the two patterns (\fig S7).

\section{Liposome size and density control pattern formation}
Fragmentation of the membrane in liposomes offered new avenues to change the physical properties of the substrate without modifying its chemical composition. Indeed, varying the lipid concentration $c$ and the size $R$ of the liposomes revealed that the properties of the dispersed substrate impact pattern formation. Starting with a suspension of liposomes with a mean radius of  {$\rm 15 \pm\, 2\,nm$ (\fig S1)} and a fixed concentration of MinD and MinE, we observed that patterns only form at intermediate lipid concentrations. As $c$ was increased, the system transitioned from a homogeneous state to spirals, coexistence, traveling waves and finally returned to a homogeneous phase (\fig\ref{fig:membraneExp}B). Interestingly, varying the lipid concentration for different mean radii, $R > 70\, \rm nm$, revealed that the range of lipid concentrations over which patterns form was size-dependent (\fig\ref{fig:membraneExp}C). Furthermore, we found that, for set concentrations of proteins and lipids, the size of the liposomes $R$ controlled the repertoire of observed patterns. Indeed, although only traveling waves were observed for large liposome suspensions ($R > 70\, \rm nm$, \fig\ref{fig:membraneExp}A and C), the spiral and coexistence phases could be restored by decreasing the MinD/E protein concentrations ratio (\fig S9). 

\begin{figure*}[t]
    \includegraphics[width=1\linewidth]{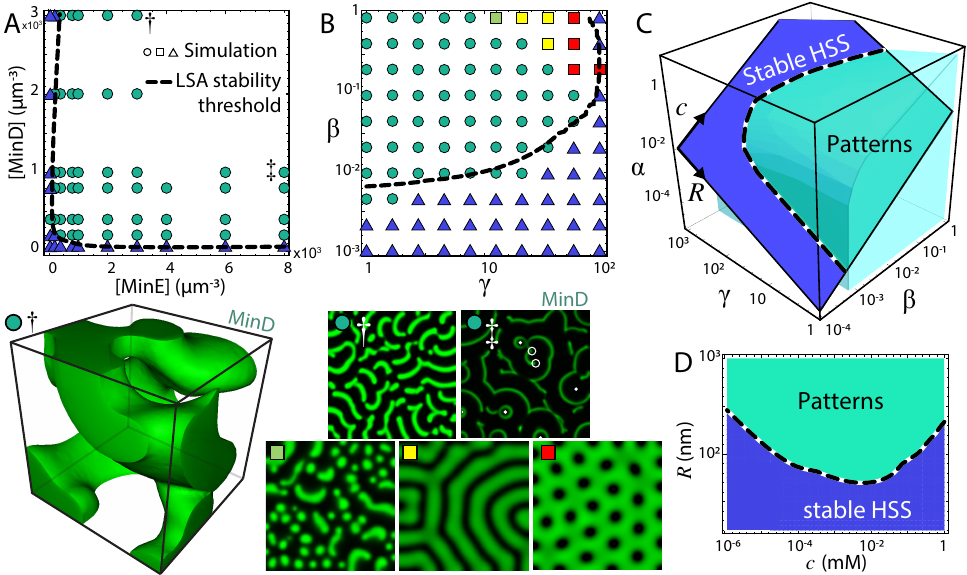}
    \caption{\textbf{Model predictions depend on membrane properties and protein concentrations.} \textbf{(A)} ([MinE],[MinD]) phase diagram with $\beta = 1$, $\gamma = 1$, and $\alpha = 1$ $\mu\text{m}{^{-1}}$ shows the dominance of wave states (teal circles) with Homogeneous Stationary States (HSS, blue triangles) occupying either low [MinE] or low [MinD] compositions; insets show two typical wave patterns, ($\dag$) plane wave-dominated and ($\ddag$) a state populated with targets (closed white point) and spirals (open white points). \textbf{(B)} ($\gamma$,$\beta$) phase diagram shows the emergence of various stationary patterns (red, yellow, and orange squares) and mixed-mode states (green squares) for ${\text{[MinE]}} = 800,{\text{[MinD]}}=1000$ and $\alpha = 1$ $\mu\text{m}{^{-1}}$. In both \textbf{(A)} and \textbf{(B)}, the dashed black lines correspond to linear stability analysis (LSA) prediction. \textbf{(C)} LSA predictions (patterns form in the teal volume) as a function of all three control parameters $\alpha,\beta$ and $\gamma$. The map between these three control parameters and the liposome size $R$ and total lipid concentration $c$, $\{\alpha\left(c\right),\beta\left(R,c\right),\gamma\left(R\right)\}$, is shown as the embedded surface with colors indicating stability. The intersection of the unstable region and this surface is shown in \textbf{(D)}, where it is parameterized by $R$ and $c$ in experimental units. For all 2D simulation images, the domain width is $50\, \rm \mu m$; for 3D, $11 \rm\mu m$.}
    \label{fig:Model}
\end{figure*}

In addition, we found that the average inter-liposome spacing $d$, which depends on the lipid concentration $c$ and liposome size $R$ (see Methods section \ref{section:MethodsExp}), governs a sharp transition at a critical cutoff spacing $d^*\sim1.5\, \mu m$. When the inter-liposome spacing was larger than this threshold,  {spatially extended} patterns did not form, irrespective of the liposome radius $R$ (\fig\ref{fig:membraneExp}D). However, in that regime, individual liposomes still displayed periodic temporal oscillations (\fig S10,  {Movie S6}), demonstrating that the dispersed Min system maintains the ability to oscillate under dilute conditions but that those oscillations cannot synchronize to form extended patterns.  {The existence of this critical inter-liposome spacing can be rationalized by considering published experimental and theoretical work on the MinDE oscillations~\cite{MinEswitch_Frey2018, MembraneBulkCoupling_Frey2021,Wurthner2022}. Recently detached MinE can exist in a transient active state and a latent conformation with a lower affinity for membrane-bound MinD (\fig\ref{fig:overview}D). The characteristic length scale associated with the active MinE can be estimated as $l_{E}=\sqrt{\frac{D_c}{\mu}} \sim 0.8\, \mu\text{m}$, where $D_{\text{c}} = 60\,\mu \rm m^2.s^{-1}$ is the diffusivity of unbound MinE, and $\rm \mu = 100\,s^{-1}$ is the active-to-latent conformational switch rate~\cite{MinEswitch_Frey2018}. If the inter-liposome spacing $d$ is smaller than $l_{E}$, cytosolic MinE is present mainly in its active conformation and can easily re-bind to a neighboring liposome. If the inter-liposome spacing $d$ is larger than $l_{E}$, cytosolic MinE goes back to its lower reactive state before finding another liposome. In that regime, the liposome oscillations are not synchronized and no spatially extended pattern will form (\fig S10, Movie S6).}

These findings highlight the complex interplay between liposome size, lipid concentration, and MinD/E concentrations in regulating both the existence and the nature of the patterns. To further disentangle the impact of the various properties of the dispersed membrane on regulating pattern formation, we next introduce a coarse-grained reaction-diffusion model for the protein-membrane suspension.

\section{A 3D reaction-diffusion model reveals that liposome diffusivity and density control pattern formation.}

To identify what characteristics of the dispersed membrane regulate pattern formation, we modified a well-established model for the MinDE system for 2D continuous membranes coupled to a 3D cytosol~\cite{MinEswitch_Frey2018}. In our revised model, we considered a homogeneous medium in which bound and unbound proteins diffuse and react in 3D (\eqns\ref{eq:model_uDD}-\ref{eq:model_ue}). The four main features of our model are: \emph{i)} we coarse-grained the liposomes and assumed that the membrane is homogeneously distributed, \emph{ii)} we converted all the protein-bound surface concentrations into volumetric concentrations, \emph{iii)} we replaced all the boundary flux terms for proteins binding to the membrane with volumetric sources and sinks, and \emph{iv)} we allowed for increased diffusivity of all membrane-bound species to account for their transport being dominated by liposome mobility and not their local on-membrane diffusivity. We thus introduced three additional parameters: $\alpha,\beta$ and $\gamma$. $\alpha$ [units: $\mu$m$^{-1}$] is the surface density of the membrane and naturally arises from converting surface concentrations into volumetric ones. The dimensionless on-rate reduction factor $\beta 
\in (0,1]$ phenomenologically accounts for the probability of co-localizing an unbound protein with the coarse-grained membrane. This is not required when the membrane is continuous and fluxes are explicitly modeled through boundary conditions as in~\cite{MinEswitch_Frey2018} but was needed here because each volume element is only partially occupied by lipid binding sites. Finally, the dimensionless diffusivity scaling factor $\gamma > 1$, modulates the diffusion coefficient of all membrane-bound proteins and complexes.

The model and its free parameters provide a framework for interpreting our experimental results. Importantly, parameters $\alpha, \beta$ and $\gamma$ depend non-trivially on the size and the density of the liposomes. We first present finite-element simulation results in which we directly tuned $\alpha, \beta$, and $\gamma $, and then outline how to connect these parameters to the experimental variables.

To develop a baseline understanding of the model’s dynamics, we first identified the pattern-forming region of phase space by systematically varying the total concentrations of MinD and MinE, and comparing the phase diagram to the phase boundaries obtained from a linear stability analysis (see Appendix \ref{sec:LSA}). Patterns appear above a critical concentration of MinD and MinE; the phase boundary measured from the simulations overlaps well with the prediction of the linear stability analysis (dashed line in \fig\ref{fig:Model}A). The morphological and dynamical character of the patterns change with the total protein concentrations. At low diffusivity ($\gamma\,=\,1$), traveling waves dominate the pattern-forming region (\fig\ref{fig:Model}A $\dag$, \fig\ref{fig:Model}A, Movie S9). These traveling waves are reminiscent of the experimental pattens (\fig\ref{fig:patterns}B): they formed domains with uniform velocity separated by lines of discontinuities. Having an excess of MinE over MinD induced qualitative changes to the pattern, and target and spirals were readily observed (\fig\ref{fig:Model}A $\ddag$, Movie S10).  {Finally, we emphasize the critical role of the active-to-latent transition of cytosolic MinE in achieving robust pattern formation. In its absence, the system becomes sensitive to variations in protein concentrations (Fig. S15), consistent with previous observations on continuous membranes \cite{denk2018a} and \textit{in vivo} \cite{Wavelength2D}. Moreover, within the pattern-forming regime, the resulting patterns deviate qualitatively from those observed experimentally (Fig. S15), underscoring that cytosolic MinE transition dynamics are essential for accurately capturing the behavior of the dispersed Min system.}

Next, we explored the impact of the membrane-dependent parameters $\alpha$, $\beta$ and $\gamma$ on pattern formation using a combination of simulations and linear stability analysis. 
Increasing the diffusivity factor $\gamma$ inhibits pattern formation, while increasing the binding rate factor $\beta$ has the opposite effect, destabilizing the homogeneous medium and promoting pattern formation (\fig\ref{fig:Model}B). Interestingly, for $\beta\sim 0.1-1$ as $\gamma$ increased, Min proteins formed various stationary Turing patterns, including stripes (Movie S11), honeycombs (Movie S12), and, for other concentrations of MinE and MinD, spots (Movie S13). At the transition between stationary and dynamic patterns, we observed mixed states, with quasi-stationary spots occupying the interface between regions exhibiting coherent waves (\fig S12 and Movie S14). The stability threshold in \fig\ref{fig:Model}B depends on the membrane surface density $\alpha$ as well. \fig\ref{fig:Model}C summarizes the full phase space predicted from linear stability analysis, with the volume enclosed by the teal surface demarcating the pattern-forming region. It shows that increasing the membrane surface density $\alpha$ counter-balances the destabilizing impact of $\beta$ leading to another phase space domain with homogeneous steady state. 

Equipped with these observations, we next endeavored to qualitatively compare experiments and theory by relating the model parameters $\alpha$, $\beta$, $\gamma$ to the experimental control parameters: the total lipid concentration $c$ and the mean liposome size $R$. First, geometric arguments reveal that the membrane surface density $\alpha$ is proportional to the lipid concentration and independent of $R$.  Second, the Stokes-Einstein relation shows that $\gamma$ is inversely proportional to $R$ and is independent of $c$. Finally, we posit that the binding rate reduction factor $\beta$ follows the law of mass action and scales with liposome number density through $\beta \sim \mathcal{N}$. Thus, as the liposome suspension gets more dilute, the effective reaction rates for the binding of bulk-diffusing proteins to the membrane decrease.  We arrive at $\beta = R\alpha \left(2/\pi\right)$. 

Using this mapping to transform our theoretical results facilitates the comparison with experiments. \fig \ref{fig:Model}C represents the map as a $\{R,c\}$-parameterized plane embedded in the $\{\alpha,\beta,\gamma\}$ simulation parameters volume. The plane intersects with the linear stability analysis-predicted phase boundary (teal surface), highlighting the experimentally accessible pattern-forming regime. The black dashed line emphasizes the stability threshold on this subspace. To further compare experiment and theory, we re-plotted the predicted phase behavior on the experimentally accessible plane (\fig \ref{fig:Model}D). Notably, for intermediate liposome size, patterns formed only for intermediary lipids concentrations. These critical concentrations depend on liposome size; increasing their radius increases the range of lipid concentrations where traveling waves form, consistent with our experimental observations (green disks in \fig\ref{fig:membraneExp}C). We can therefore conclude that the spacing between reaction sites and their mobility are key parameters that control the emergence of patterns. We next investigated the quantitative impact of these physical parameters on wavelength selection, focusing on the traveling wave regime.

 \section{Wavelength selection depends on the physical parameters of the dispersed membrane.}
\begin{figure}
    \includegraphics[width=1\linewidth]{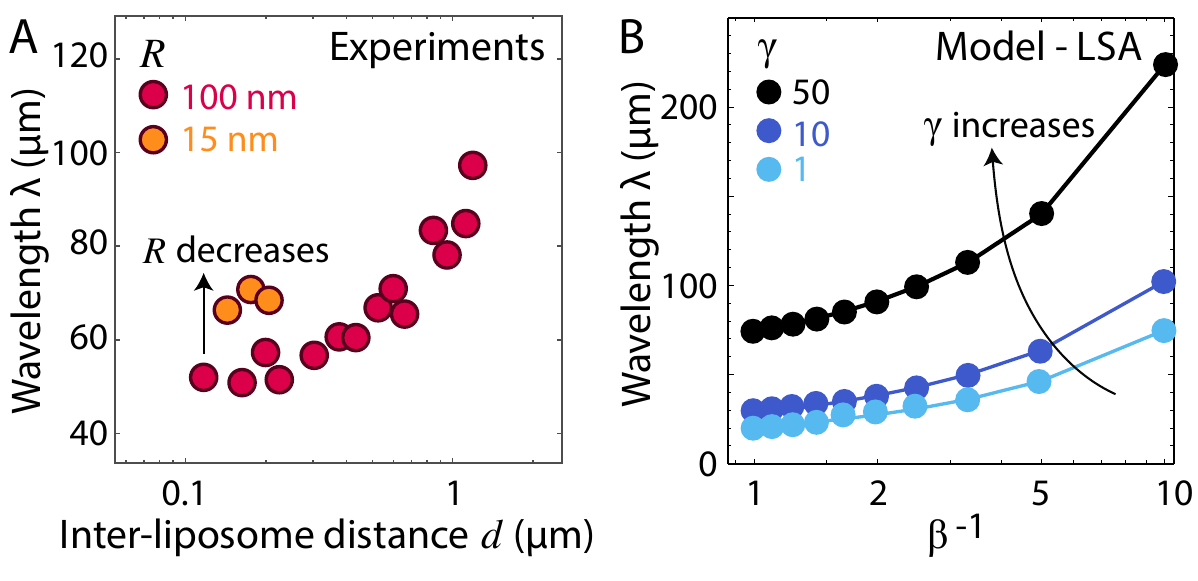}
    \caption{ \textbf{Wavelength depends on dispersed membrane properties.} \textbf{(A)} The wavelength $\lambda$ increases when the inter-liposome distance $d$ increases and decreases when the mean liposomes radius $R$ decreases. Orange and red data points respectively correspond to: $R = 15\,\rm nm$ and $R = 100\,\rm nm$. Each data point is averaged over 2-5 replicates. Error bars are smaller than the marker size. \textbf{(B)} Linear stability analysis-predicted wavelength as a function of $\beta^{-1}$ for $\gamma=\{1,10,50\}$, respectfully light blue, blue, and black.
    }
    \label{fig:Wavelength}
\end{figure}

At fixed liposome size, increasing the inter-liposome spacing by decreasing the lipid concentration induced an increase of the wavelength (red dots in \fig\ref{fig:Wavelength}A). Linear stability analysis of our model predicts a similar trend: decreasing $\beta$, which is equivalent to slowing down binding kinetics, increases the wavelength (\fig\ref{fig:Wavelength}B). These predictions are consistent with previous numerical studies on the 2D-3D system, which independently varied the MinD-to-membrane and MinD-to-bound-MinD binding rates (resp. $\kd$ and $\kdD$) ~\cite{Wavelength2D} , both of which, in our model, scale with $\beta$ (\eqns \ref{eq:model_uDD}-\ref{eq:model_ue}).

Holding the inter-liposome spacing constant while decreasing the liposome size $R$ led to an increase of the wavelength, \fig\ref{fig:Wavelength}A (orange points). These observations are consistent with our model as well, where increasing diffusivity by increasing $\gamma$ increases the wavelength (\fig\ref{fig:Wavelength}B). 

Intuition for the opposing roles of reactive and diffusional processes on wavelength selection can be garnered by considering a dimensionless reaction-diffusion lengthscale $\sim\sqrt{\gamma \beta^{-1}}$, derived from our model's parameters. Although the numerical trends deviate from this scaling, this lengthscale provides a rule of thumb for interpreting the results: a higher diffusion tends to increase both the width of the reaction fronts and the pattern wavelength, whereas higher reaction rates shorten these lengthscales. Overall, we conclude that the properties of the liposomes impact both diffusive and reactive aspects of the Min system. The coarse-grained model enables us to disentangle their multiple influences, offering insights into how physical parameters of the dispersed membrane shape phase boundaries and select the wavelength of the emerging patterns. 
 {Finally, we note that the liposome spacing and size also impact the temporal period selection in the traveling waves regime, both in the experiments and in the simulations (\fig S13).}

\section{Conclusion and Discussion}

Our study revealed how the physical properties of a dispersed membrane control the formation of dynamical protein patterns.  {Our main finding is that the Min system can robustly form three-dimensional patterns despite the complete loss of membrane continuity. Cytosolic coupling—whereby unbound active proteins diffuse through the bulk fluid before rebinding to other liposomes—is essential for enabling neighboring liposomes to synchronize their oscillations, thereby triggering the emergence of spatially extended patterns with characteristic length scales spanning hundreds to thousands of liposomes. Synchronization is lost above a critical inter-liposome distance, which is controlled by the reaction-diffusion lengthscale of the active-to-latent conversion of cytosolic MinE. Previous studies have already highlighted the importance of cytosolic coupling for Min pattern formation on continuous membranes, showing that it enables pattern formation despite membrane spatial heterogeneity, such as small two-dimensional obstacles \cite{GeoSense_Schweizer2012} or patches with preferential MinD attachment \cite{halatek2012highly}. Our work extends this principle to fully fragmented, freely diffusing membranes in three dimensions.} We conclude that while membrane continuity might be required for MinDE to perform its function of localizing MinC and FtsZ \emph{in vivo}, it is not necessary for pattern formation.

We found that changing the lipid concentration and liposome size distribution offers a novel route for controlling the effective membrane binding rates and the diffusivity of membrane-bound proteins. These alterations enable access to various dynamical 3D patterns and the modulation of their wavelength without changing the biochemical properties of the proteins. While previous studies of non-equilibrium chemical systems have established the importance of physical properties, such as phase continuity \cite{Vanag2009,Alonso2011,Tomography_BZAOT} and mobility \cite{castets1990,Lengyel1991}, on pattern formation, ours demonstrates their applicability to biomolecular systems.  {Our findings raise a broader question: to achieve a desired pattern, is it more effective to alter biochemical reaction rates or physical properties? Qualitatively, a desired spatiotemporal pattern may be achieved through multiple pathways, but from an engineering point of view, it may be possible to identify a minimal set of interventions by weighing the relative “cost” or feasibility of biochemical versus physical modifications. A more systematic exploration of how individual reaction rates influence pattern formation in the dispersed-membrane system could clarify these tradeoffs.}

Our model qualitatively predicts the criteria for pattern formation and decouples the lipids' multifaceted influence on the dynamics into reactive and diffusive contributions. In what follows, we identify limitations that may inspire future investigations. First, we note that our model predicts that patterns form under membrane conditions more dilute (low $\beta$ and low $\alpha$) than experimentally observed. We postulate that this is due to a breakdown in liposome synchronization in the experiment, which challenges the continuum assumption of our model. The lengthscale at which pattern formation is lost is commensurate with the reaction-diffusion length associated with the transition of MinE from its active to its latent conformation~\cite{MinEswitch_Frey2018}. Beyond this distance, this suggests that the slower binding of latent-form MinE is insufficient for neighboring liposomes to strongly influence one another, leading to desynchronization and subsequent loss of spatially extended patterns.

Second, we note that our model consistently predicts the formation of a variety of static Turing patterns at high diffusivity $\gamma$ and high on-rate reduction factor $\beta$. In contrast, stationary patterns were observed only marginally in our experiments (\fig  S11 and Movie S8). Interestingly, \emph{in vivo} dynamics of the Min system implanted in eukaryotic cells can also exhibit stationary spot and stripes patterns~\cite{Coyle2024}, reinforcing Min's pattern-forming versatility over a range of physical scenarios. A distinct MinE variant also forms a variety of static patterns on continuous membranes~\cite{Glock2019}. Future work should focus on testing this theoretical prediction more robustly by further reducing the liposome's mean radius, increasing their concentration, and elucidating the concentration dependence.  {Reciprocally, a number of phenomena are observed in experiments that prompt further theoretical explorations. In particular, the coexistence of multiple patterns or similar patterns with discontinuous properties (as the planar waves forming sharp singularities between domains with distinct traveling direction reported in \fig S5 and Movie S3). The establishment of these domains and the slow evolution of discontinuities are evocative of some generic singularities in oscillatory media that were observed about 50 years ago and recently classified mathematically~\cite{zaikin1970concentration,howard1977slowly,sandstede2004defects}}.

Finally, our experiments clearly show that two distinct dynamical patterns can form; one characterized by closely packed traveling waves and another by spiral waves. The contrasting dynamics between the two patterns is most apparent in the coexistence phase. By contrast, simulations exhibit a more continuous transition from packed planar waves to dynamical patterns with spirals; however, even in these cases, the spirals are more densely packed than in the experiment (see insets of \fig\ref{fig:Model}A$\dag$, $\ddag$).  {While previous work showed that a "commensurability condition" could predict the transition from chaotic dynamics to traveling waves \cite{Wurthner2022}, our analysis of the dispersion relations did not reveal a similar criterion for distinguishing between types of non-stationary patterns or for identifying transitions between stationary and non-stationary regimes. Examining the stability of fully developed spatial patterns—such as bands and spots—may offer deeper insights into the mechanisms driving these transitions \cite{Goychuk2024}}

The contrast between theory and experiment opens exciting opportunities for investigating the origin of the spiral and coexistence phases. We formulate below a few hypotheses that could be tested experimentally or theoretically. First, it is possible that some of the model parameters, including the chemical rates and diffusivity of the various proteins, can be further optimized. It may be appropriate to measure, when possible, the binding rates of Min proteins onto liposomes instead of relying on published estimates and measurements for proteins binding on a flat, continuous membrane. Such measurements would elucidate the possible impact of membrane curvature on binding rates~\cite{MILEYKOVSKAYA200322193, JOHNSON2024401}. Last, the experimentally observed spiral and coexistence phases may arise from the confluence of excitability, heterogeneity, breaking down of continuum assumptions, and noise near phase transitions. Our model, by coarse-graining the liposomes, implicitly assumes that the macroscopic behavior of a large ensemble of heterogeneous liposomes can be effectively described by a homogeneous system with an average liposome size and dynamics. However, in non-linear media, macroscopic behaviors in heterogeneous systems often differ from those in homogeneous ones~\cite{strayer2003kind,altschuler2010cellular,hermann2012heterogeneous}. In our case, heterogeneity could manifest in multiple ways: local oscillations that fail to synchronize globally yet produce complex mesoscopic patterns, persistent local pattern formation disrupted at larger scales, or local oscillations that do not translate into mesoscale structures. In addition, heterogeneity and stochastic fluctuations near bifurcations are known to drive complex pattern formation, including attractor switching and resonances ~\cite{MERON19921, Buceta2003,Touboul2012, Touboul2020, Mittal2022}. Given that our experimental system is inherently discontinuous and polydisperse (\fig S1), the highly dynamic spiral and coexistence phases may reflect the interplay of non-linear excitability, heterogeneity, and stochastic fluctuations. Future work should focus on distinguishing between these scenarios to unravel the precise mechanisms underlying these emergent patterns.

Our findings have broad implications for the physical underpinnings of biology and design of non-equilibrium bioinspired materials. The principles applied here, in particular, the binding of proteins to a motile substrate with a tunable diffusivity, are generic and relevant for protein systems that react on a substrate. Given the rapid advancements of synthetic biology and \emph{de novo} protein design, such physical considerations will be important when engineering systems based on non-equilibrium biomolecular constituents.\\

{\bf Acknowledgments.} We thank Pr. Petra Schwille and Dr. Michaela Schaper for the His-MinE plasmid.  We thank Dr. Shibani Dalal, Director of the Brandeis Biomaterial Facility for help with protein purification and labeling. We thank Dr. Joseph Lopes, Lance Babcock, and Dr. Bibi Najma for initial work on the 2D Min system. We thank Dr. Wei-Shao Wei for help with the dynamic light scattering measurements. This work was supported by the Brandeis NSF Materials Research Science and Engineering Center (MRSEC) Grant No. DMR-2011846. G.D. acknowledges support from NSF CAREER Grant No. DMR-2047119. We also acknowledge the use of the optical and biomaterial facilities supported by NSF MRSEC Grant No. DMR-2011846. Development of the computational model and analysis was supported by
DOE BES No. DE-SC0022280 (M.M.N.). \\

\textbf{Author contributions:} 
A.C. and G.D. designed research; A.C. and M.M.N. performed
research; A.C., M.M.N., J.T., and G.D. contributed new reagents/analytic tools; A.C. and
M.M.N. analyzed data; and A.C., M.M.N., J.T., and G.D. wrote the paper.\\

\appendix
\section{Theory}\label{section:MethodsModel}

\subsection{PDE model}
We adapt a PDE model of MinDE dynamics that tracks the spatiotemporal dynamics of various forms of membrane-bound and freely-diffusing ``bulk'' MinD, MinE, and MinD-MinE complexes~\cite{MinEswitch_Frey2018}. The model accounts for the interrelated dynamics of 7 species in total, 4 in the cytosol and 3 on the membrane. Beginning in the cytosol, we have two forms of MinD, those bound to ATP $\uDT$ and those bound to ADP $\uDD$, and two forms of MinE, a ``reactive'' form  $\uEr$ that binds rapidly to membrane-bound MinD, and a slower-reacting ``latent'' form  $\uEl$. On the membrane, MinD $\ud$, MinE, and MinD-MinE $\ude$ complexes are tracked. Following the convention from Refs.~\cite{MinEswitch_Frey2018,Model_Frey2023} and others, capital subscripts denote cytosolic species, where lowercase indicate membrane-bound. The units of these species are inherited from the dimension of their domain: all bulk fluid species are in units of [\#/ $\mu$m${^3}$] with membrane-bound in [\#/ $\mu$m${^2}$]. We begin by restating the  2D membrane and 3D bulk equations with appropriate reactive bound conditions coupling them:

\begin{widetext}
\begin{align}
    \partial_t \uDD &= \Dc \nabla^2 \uDD -\lambda_{\text{DD}} \uDD \\
    \partial_t \uDT &= \Dc \nabla^2 \uDT + \lambda_{\text{DD}} \uDD\\
    \partial_t \uE^r &= \Dc \nabla^2 \uE^r - \mu \uE^r \\
    \partial_t \uE^l &= \Dc \nabla^2 \uE + \mu \uE^r\\
    \partial_t \ud &= \Dd \nabla^2 \ud 
    +\left(\kD +\kdD \ud \right) \left.\uDT\right|_{\partial\Omega}
    - \ud
    \left(
\left.\uE^r\right|_{\partial\Omega} \kdE^r +
\left.\uE^l\right|_{\partial\Omega} \kdE^l
\right)
    -\ked \ud\ue\\
    \partial_t \ude &= \Dde \nabla^2 \ude 
    +\ud
    \left(
\left.\uE^r\right|_{\partial\Omega} \kdE^r +
\left.\uE^l\right|_{\partial\Omega} \kdE^l
\right)
    -\kde\ude +\ked \ud\ue \\
    \partial_t \ue &= \De \nabla^2 \ue
     +\kde \ude -\ked \ud\ue -\ke\ue   
\end{align}

These 2D and 3D PDE's are coupled through the following boundary fluxes $f_{\partial\Omega}$:

\begin{align}
    \Dc \left.\bn\cdot\nabla\uDD\right|_{\partial\Omega} &= \kde\ude \\
    \Dc \left.\bn\cdot\nabla\uDT\right|_{\partial\Omega} &= -\left(\kD + \kdD\ud \right)\uDT \\
    \Dc \left.\bn\cdot\nabla\uE^r\right|_{\partial\Omega} &= -\kdE^r \ud\uE^r + \ke \ue \\
    \Dc \left.\bn\cdot\nabla\uE^l\right|_{\partial\Omega} &= -\kdE^l \ud\uE^l
\end{align}
\end{widetext}

Given the separation of length scales between the size of liposomes $\sim\mathcal{O}\left(10-100\right)$ [nm] and that of typical patterns $\sim \mathcal{O}\left(100\right)$ [$\mu$m], we develop a continuum model by coarse-graining away the fine features of the liposomes. Each volume element in our model contains an concentration of lipids $c$ arranged into liposomes of size $R$ and a mix of free and membrane-bound proteins but disregards detailed dynamics on and between liposomes.

To begin, we first transform the membrane concentrations with units [\#/$m^2$] to volumetric concentrations. We consider a volume element with volume $V$ containing lipid with a surface area $A$. The total number of bound proteins is then $N=Au$. Dividing by the volume gives the volumetric concentration $u^*=\frac{A}{V} u \equiv \alpha u$ or $u = \alpha^{-1} u^*$. This conversion will be applied to all surface-bound species $\{\ud,\ude,\ue\}$. Similarly, the surface fluxes $f_{\partial\Omega}$ \eqns A8-A10 can be converted to volumetric source terms $f_{\Omega}$ in \eqns A1-A3 by assuming that such fluxes act over the same effective area $A$ and then dividing by $V$ to find the rate of change per time per volume. Thus $f_{\Omega}=\frac{A}{V}f_{\partial\Omega}\equiv\alpha f_{\partial\Omega}$.

Additionally, we note that in \eqns 8-10, $\uDT$ and $\uE$ represent the activated MinD and MinE concentrations \emph{at} the membrane. In our dispersed membrane model, a given volume element contains a concentration of lipid upon which binding can occur. To account for the impact of liposome availability, we scale all bulk-to-membrane processes with $\beta\in(0,1]$. Finally, we let the membrane-bound diffusivities be increased by a factor $\gamma>1$ to account for the mobility of the liposomes to which these species are bound.

Combining these facts and simplifying gives the following bulk equations where all concentrations are in [\#/$m^3$]:

\begin{widetext}
\begin{align}
    \partial_t \uDD &= \Dc \nabla^2 \uDD -\lambda_{\text{DD}} \uDD
    + \kde\ude^*,
    \label{eq:model_uDD}
    \\
    \partial_t \uDT &= \Dc \nabla^2 \uDT + \lambda_{\text{DD}} \uDD
     -\left(\beta\alpha\kD + \beta\kdD\ud^* \right)g\left(\ud,\ude\right)\uDT,
    \label{eq:model_uDT}
    \\
    \partial_t \uE^r &= \Dc \nabla^2 \uE^r
    -\beta\kdE^r \ud^*\uE^r + \ke\ue^* -\mu \uE^r,
    \label{eq:model_uEr}
    \\
    \partial_t \uE^l &= \Dc \nabla^2 \uE^l
    -\beta\kdE^l \ud^*\uE^l +\mu \uE^l,
    \label{eq:model_uEl}
    \\
    \partial_t \ud^* &= \gamma\Dd \nabla^2 \ud^* 
    +\left(\beta\alpha\kD +\beta\kdD \ud^* \right)g\left(\ud,\ude\right)\uDT
    -\beta\ud^* \left(\kdE^r\uE^r +\kdE^l\uE^l \right) 
    -\frac{1}{\alpha}\ked \ud^*\ue^*,
    \label{eq:model_ud}
    \\
   \partial_t \ude^* &= \gamma\Dde \nabla^2 \ude^* 
    +\beta\ud^* \left(\kdE^r\uE^r +\kdE^l\uE^l \right) 
    -\kde\ude^*
    +\frac{1}{\alpha}\ked \ud^*\ue^*, 
    \label{eq:model_ude}
    \\
    \partial_t \ue^* &= \gamma\De \nabla^2 \ue^*
     +\kde \ude^*
     -\frac{1}{\alpha}\ked \ud^*\ue^*
     -\ke\ue^*,
     \label{eq:model_ue}
\end{align}
where we have also added a factor
\begin{equation}
g\left(\ud,\ude\right)=\left(\frac{u_{\text{d,max}}\alpha -\ud^*-\ude^*}{u_{\text{d,max}}\alpha }\right)
\label{eq:surfacemax}
\end{equation}
that prevents the total concentration of minD from exceeding a prescribed value. We found this factor increases the numerical stability, allowing us to explore a larger range of parameters $\{\alpha,\beta,\gamma\}$.

\begin{table}[h]
\begin{center}
\begin{tabular}{ |c|c|c| } 
 \hline 
parameter description & symbol & value \\
 \hline 
binding rate of bulk minD to membrane & $\kD$ & 0.0625 \\
cooperative binding rate of MinD to membrane-bound MinD & $\kdD$ & 0.02 \\
rate of reactive-MinE recruitment from bulk to membrane-bound MinD & $\kdEr$ & 0.2 \\
rate of latent-MinE recruitment from bulk to membrane-bound MinD & $\kdEl$ & 0.002\\
rate of MinDE complex degradation & $\kde$ & 0.34 \\
rate of complex formation between membrane bound MinE and MinD & $\ked$ & 0.01 \\
membrane-bound MinE off rate & $\ke$ & 0.01 \\
MinD recycling rate: ADP$\rightarrow$ ATP & $\lamDD$ & 6 \\
bulk MinE inactivation rate & $\mu$ & 100 \\
maximum surface concentration of MinD & $u_{\text{d,max}}$ & 5e3 [\# /$\mu$m$^{-2}$] \\
\hline
average MinD concentration & $\nD$ &  $\mathcal{O}\left(10^2-10^4\right)$  [\# /$\mu$m$^{-3}$]\\
average MinE concentration & $\nE$ &  $\mathcal{O}\left(10^2-10^4\right)$ [\# /$\mu$m$^{-3}$] \\
\hline
diffusion coefficient for unbound MinD,MinE $\{\uDT,\uDD,\uEr,\uE\}$ & $D_{\text{c}}$ & 60 \\
diffusion coefficient of bound MinD ($\ud$) & $D_{\text{d}}$& 0.013\\
diffusion coefficient of bound MinE ($\ue$) &$D_{\text{e}}$ & 0.005\\
diffusion coefficient of bound MinD-MinE ($\ude$) & $D_{\text{de}}$& 0.013\\
\hline
lipid surface area to volume ratio & $\alpha$ & $10^{-3}-10^1$ [$\mu$m$^{-1}$] \\
bulk-to-membrane reaction rate reduction factor & $\beta$ & $10^{-3}-10^0$  [-]\\
diffusivity factor for membrane-bound proteins & $\gamma$  &  $10^0-10^3$[-] \\
\hline
\end{tabular}
\end{center}
\caption{Table of model parameters \label{table:params}}
\end{table}

\end{widetext}

\subsection{Linear Stability Analysis \label{sec:LSA}}
To make analytical predictions of pattern-forming conditions, we perform a linear stability analysis (LSA) on our model \eqns \ref{eq:model_uDD}-\ref{eq:model_ue} by examining the growth of perturbations around a homogeneous steady state. The linearized dynamics are given by
\begin{equation}
\partial_t\deltabu = \bD \nabla^2 \deltabu + \left.\mathcal{J}\right|_{\bu=\bu^*}\deltabu,
\end{equation}
where $\bD$ is a 7$\times$7 diagonal matrix of diffusivities and $\mathcal{J} = \frac{\partial\mathbf{G}}{\partial\bu}$ is the Jacobian matrix of the chemical dynamics $\mathbf{G}\left(\bu\right)$ evaluated at a fixed point satisfying $\mathbf{G}\left(\bu^*\right)=0$. We note that the inclusion of the factor \eqn\ref{eq:surfacemax} to \eqns \ref{eq:model_ud} and \ref{eq:model_uDD} allows $\mathbf{G}$ to support multiple, physical fixed points. For simplicity, we remove the pre-factor when performing the stability analysis. In this case, there is only one physical solution, $\bu^* \in \mathbb{R}+$. We find that this simplification does not impact the overall agreement between the linear stability analysis and numerical predictions. We let the form of the perturbation to the homogeneous steady state be
\begin{equation}
    \deltabu \sim \exp{\left(\sigma t + ikx\right)},
\end{equation}
where $\sigma\in \mathbb{R}$ is the growth rate and $k\in \mathbb{R}$ is the wavenumber, where we've assumed Cartesian coordinates. 
\begin{equation}
    \sigma \mathcal{I} = -\bD k^2 + \mathcal{J}
\end{equation}
Solving for the values of $\sigma$ that satisfy $\det\left(-\bD k^2 + \mathcal{J}-\sigma \mathcal{I}\right)=0$. Doing so yields the dispersion relation $\sigma\left(k\right)$, which we examine for regions where $\sigma > 0$ as a function of the various model parameters, see \fig\ref{fig:Model}A-D. In \fig\ref{fig:Wavelength}B, we also plot the fastest growing mode's wavelength $\lambda = 2\pi/k^*$, $k^* \in \arg\max\limits_{k} \sigma\left(k\right)$.  {We plot dispersion relations in \fig S14 for parameters corresponding to each of the simulation movies S9-S14. Qualitatively, all display the same features: a band of unstable wavenumbers bounded away from zero and a region at low wavenumber in which slowly growing modes possess an oscillatory component. Thus, the coarse features at the level of linear stability do not immediately highlight contrast between the various stationary and non-stationary patterns.}

\subsection{Numerical Methods}
We solve the coupled set of nonlinear partial differential equations \ref{eq:model_uDD}-\ref{eq:model_ue} using the finite element software COMSOL \cite{mingit} with parameters listed in Table 1. For the majority of simulations, we solve for the dynamics on a two-dimensional 50 [$\mu$m] x 50 [$\mu$m] domain with no-flux boundary conditions over a time interval $\mathcal{O}(10^3)$ [s] using a relative tolerance of $10^{-5}$. Simulations were conducted on a high-performance desktop with a 24-physical core Intel Xeon 3.0 GHz processor and 96 GB of error correcting code DDR4 RAM.

\section{Experimental methods}\label{section:MethodsExp}

\subsection{Samples preparation}
\subsubsection{Proteins purification}
His-MinD and His-MinE were purified using the plasmids pET28a-His-MinD-MinE, pET28a-eGFP-MinD  and pET28a-MinE, respectively. The MinD plasmids were purchased from Addgene (\url{https://www.addgene.org/Petra_Schwille/}) and the His-MinE plasmid was gifted from the Schwille lab. In short, proteins were expressed in \textit{E.Coli} BL21 pLysS and purified through a His-Trap Ni-NTA affinity column, following the detailed protocol published in~\cite{ProtocolSchwille}. Purified proteins were further size-excluded in a storage buffer (50.0 mM HEPES/KOH at pH 7.2, 150.0 mM KCl, 10$\%$ glycerol, 0.1 mM EDTA and 0.4 mM tris(2-carboxyethyl)phosphine). Proteins were flash frozen in liquid nitrogen and stored in small aliquots at $\rm -80 \degree C$. 

\subsubsection{Protein labeling}
The labeling of His-MinE with CF-568 maleimide (Biotum 92024) was performed according to the dye manufacturer’s instructions. In brief, the protein stock solution was mixed with 3-fold molar excess of reactive dye. The reaction mixture was incubated overnight at $4\degree \rm C$,  protected from light. A spin desalting column (Thermo Fisher Scientific, Zeba 7K MWCO, 0.5ml, Prod $\#$89882) was used to remove most of the unbound dye.

\subsubsection{Measure of proteins concentration}
Protein concentrations were determined using a linearized Bradford assay with Bovine Serum Albumin as a reference. We used a NanoDrop Spectrophotometer (Thermo Scientific™ NanoDrop™ 2000c) and repeated measurements at least three times for each concentration. 

\subsubsection{Liposomes preparation}
We prepare a solution of $\rm 70\,mol\,\%$ DOPC (Avanti Polar Lipids, 18:1 ($\Delta$9-cis) PC; SKU, 850375C) and $\rm 30\, mol\,\%$ DOPG (Avanti Polar Lipids, 18:1 ($\Delta$ 9-cis) PG, SKU: 840475C), dissolved in chloroform.  {This lipid composition is known to allow efficient MinDE membrane binding and unbinding and pattern formation \textit{in vitro}~\cite{ProtocolSchwille, Vecchiarelli2014}}. Chloroform is then evaporated under a vacuum pump for three hours. The dry lipid film is rehydrated in Min buffer (25 mM Tris-HCl pH 7.5, 150 mM KCl, 5 mM $\rm MgCl_{2}$) to a final lipid stock concentration of $5\, \rm mM$. 

$\rm 15\,nm$ liposomes are prepared using micro-tip sonication (OR250-220 Omni Ruptor 250 Ultrasonic Homogenizer) at middle range power until clarity ($\sim 3\,$min).
Larger liposomes ($\rm 70\,nm$, $\rm 100\,nm$, $\rm 110\,nm$) are prepared using an extruder (Avanti Polar Lipids, Mini Extruder, SKU: 610000-1EA) with appropriate porous membranes (Whatman® Nuclepore™, Track-Etched Membranes, polycarbonate filter). Liposome solutions are stored at $\rm 4 \degree C$ for a maximum of two weeks.

Labeled liposomes are prepared using the same protocol, starting with a mixture of $\rm 69\,mol\,\%$ DOPC (Avanti Polar Lipids, 18:1 ($\Delta$9-cis) PC; SKU, 850375P), $\rm 30\,mol\,\%$ DOPG (Avanti Polar Lipids, 18:1 ($\Delta$9-cis) PG, SKU: 840475P) and $\rm 1\,mol\,\%$ Liss Rhod-PE (Avanti Polar Lipids, 18:1, SKU: 810150C). 

Size distribution, concentration and diffusion coefficient of each solution of liposomes were measured using Dynamic Light Scattering (Wyatt Technology DynaPro™
NanoStar™). The stock solution was diluted 1000$\times$ in Min buffer. Measurements were made at $\rm 25\degree C$, using a $\rm 3\, \mu L$ quartz-cuvette. We confirmed that the liposome suspensions were stable over the two weeks time period of liposomes storage.

\subsubsection{Sample preparation}
In order to ensure time stability of our samples under image acquisition, we prepared our active mixture by diluting our stock proteins to the working concentration in Min buffer supplemented with: ATP ($\rm 2.5\,  {m} M$),  ATP regenerating system containing 31 mM Phosphoenolpyruvic acid monopotassium salt (PEP, Beantown Chemical, 129745) and pyruvate kinase and lactate dehydrogenase enzymes 4.5$\%$ v/v (PK/LDH, Sigma, P-0294), antioxidants to reduce photobleaching (glucose $\rm 20.3\, mM$, DTT $6\, \rm  mM$, glucose oxidase $\rm 240\, \mu g/mL$, catalase $\rm 43\, \mu g/mL$, and Trolox $\rm 0.3\, \mu M$), Bovine Serum Albumine (BSA, Sigma, $\rm 6.6\, mg/mL$), and Polyethylene glycol (PEG 35k Fluka, $\rm 0.8\, \%\ w/v$).

\subsubsection{Microfluidic channels and sample loading}
Our active mix was confined in rectangular flow channels. Our microfluidic device was assembled using two glass slides spaced by a layer of Parafilm 'M' (height $\rm \sim 100\, \mu m$). The glass surfaces were coated with a PEG-acrylamide brush to prevent protein or lipid adsorption. Parafilm spacers were cut following an homemade design on a cutting machine (CAMEO, Silhouette 2). They were then placed between the two glass surfaces and slightly softened at $\rm 65\, \degree C$ to ensure good adhesion to the glass surfaces. 

Experiments performed with channels at different heights used layers of double-sided tape (3M™ Double Coated Tape 9629PC) instead of parafilm spacers. 

Samples of $\rm 8\, \mu L$ were loaded into each channel by capillarity using a micropipette. Finally, channels were sealed using a UV-curing optical adhesive (NOA 81, Norland Products Inc.) to ensure no external flow during the time of acquisition of the experiment.

\subsection{Microscopy}
\subsubsection{Confocal microscopy}
3D visualisation of the patterns was obtained by volume reconstruction from z-slices images taken by confocal microscopy (Nikon AX-R), using 10X or 20X objectives. Samples were illuminated using the appropriate wavelength ($\rm 488\,nm$ for MinD-gfp and $\rm 561\,nm$ for CF568-MinE or Liss Rhod-PE lipids) from the light source.
The step size was set to $\rm 3.5\, \mu m$ and the time interval between two images to $524\,\rm ms$, both ensuring sufficient resolution in time and space to avoid loss of information on the dynamics and the structure, without introducing any bias between two z slices.  

\subsubsection{Widefield fluorescence microscopy}
Time-lapse image acquisitions were taken using a Nikon Eclipse Ti2 inverted fluorescence microscope, equipped with a LED light source (Lumencor sola 80-10244). The multiposition data acquisitions, the time interval between the images, exposure time,
and illumination intensity were controlled by a micromanager software~\cite{mumanager2014}.
The typical delay between two successive images of the same field of view was set to be $\rm 30\,s$. 
A Hamamatsu camera (ORCA4.0V1, pixel size $\rm 6.45\, \mu m$) and a 4X or 10X objective (Nikon Plan Apo) were used for imaging Min proteins concentration field.

Note that, for the ease of data manipulation, we only imaged middle plane z-slices timelapses instead of full 3D volumes to construct our phase diagrams. We also restricted our study to $100\, \rm \mu m$ channels in order to limit temporal reorganization in z, as patterns tend to be mostly homogenous in z when the system is confined ($H \le \lambda \sim 100\, \rm \mu m$), \fig S2A. We verified that the channel height is not a parameter impacting wavelength selection in our system, \fig S2B.


\subsection{Image and data analysis}
\textbf{Patterns characterization:} The type of patterns for each experiment is determined manually, following the criteria given in the main text (Section~\ref{section:patterns} and \fig\ref{fig:patterns}). \\ 

\textbf{Kymographs and intensity profiles:} We used ImageJ to reshape our images as space-time kymographs and extract the intensity profile values. The intensity $I$ was normalized as: $ (I \ - <I>)/<I>$, where $<.>$ denotes the spatial average. \\

\textbf{Wavelength and period:} We measured the wavelength and the period by using a homemade automatized MATLAB code. Briefly, we reshaped intensity images into space-time kymographs along x or y slices. We then performed spatial (resp. temporal) autocorrelation function and detected the wavelength (resp. period) as the $1^{st}$ maxima. The wavelength and the period that we report for each experiment are the average values over 25 x-slices and 25 y-slices. \\

\textbf{Phase field and defects:} To extract the positions of the core of the spirals, we measured the local winding number in the phase field. The phase field $\Phi (x,t)$ is computed following the protocol detailed in~\cite{TurbulenceDefects_Fakhri2020}. Clockwise (resp. anti-clockwise) spirals correspond to a winding number equals to +1 (resp. -1).  \\

\textbf{Orientation field:} To compute the orientation field $\theta (x,y,t)$, we measured the angle of the velocity field of the waves. The velocity field corresponds to the spatial 2D gradient of the phase field, such that $\theta (x,y,t) = tan^{-1}\left(\nabla(\Phi (x,y,t))\right)$. \\

\textbf{Relationship between liposome properties and model parameters:}
Here we derive, in detail, the map between the three model tuning parameters $\{\alpha,\beta,\gamma\}$ and the two experiment parameters governing the lipid dispersion $\{R, c\}$, respectively, the liposome radius and the lipid concentration. We begin with the membrane area density or surface-to-volume ratio, which follows directly from geometric considerations such that
\begin{equation}
\alpha = \frac{A}{V} = 4\pi R^2\mathcal{N},
\end{equation}
where we've assumed liposome monodispersity. To relate the number of liposomes $\mathcal{N}$ to the number of lipids, we use the geometric relation
\begin{equation}
   \mathcal{N}=
    \frac{n}{V}\frac{a}{4\pi\left[R^2+\left(R-h\right)^2\right]},
\end{equation}
where $n$ is the number of lipids contained per individual liposome, $a=0.7\, \rm nm^2$ is the area of a single lipid head, and $h \sim 5\, \rm nm$ is the thickness of the lipid membrane \cite{FractionLipids_2021}. The number density $n/V$ can then be cast in terms of a concentration $c$ using Avogadro's number $\mathcal{N}_{\text{A}}$ such that the final expression for $\alpha$ becomes
\begin{equation}
\alpha = \frac{a R^2 c \mathcal{N}_{\text{A}}}{\left[R^2+\left(R-h\right)^2\right]},
\end{equation}
We simplify the expression by noting that $h \ll R$ for most cases, giving
\begin{equation}
\alpha = \frac{1}{2} a c {N}_{\text{A}}.
\label{eqn:alpha}
\end{equation}
For the smallest radii we consider, the error of this approximation is less than 10\%.
Thus, to leading order, the membrane density is independent of liposome geometry and only depends on the concentration of lipids. 

We now consider the on-rate reduction factor $\beta$ by assuming that reaction rates that capture proteins finding liposomes in solution will vary in proportion to the density of those liposomes, such that $\beta \sim \mathcal{N}$. To make this expression dimensionless, we choose a prefactor $1/\mathcal{N}_0$ such that $\beta = \mathcal{N}/\mathcal{N}_0 \rightarrow 1$ (its maximum physical value) as the average center-to-center distance $w$ between liposomes, defined as $w = \left(\frac{\mathcal{N}}{V}\right)^{-\frac{1}{3}}$,
 approaches $2R$, such that
\begin{equation}
    2R = \left(\frac{\mathcal{N}_0}{V}\right)^{-\frac{1}{3}}.
\end{equation}
Bringing together these elements gives the expression for $\beta$ 
 \begin{equation}
     \beta = \frac{1}{\pi}c R a \mathcal{N}_{\text{A}}  = \frac{2}{\pi}\alpha R,
     \label{eqn:beta}
 \end{equation}
 where we utilized \eqn B2 and \eqn B3, and again made use of $ h \ll R$. To interpret our experimental phase diagrams, we also utilize the average surface-to-surface distance between liposomes $d$ defined as 
 \begin{equation}
     d = w - 2R.
\label{eqn:liposomedistance}
 \end{equation}

Finally, to relate $\gamma$ to experimental control parameters, we utilize the Stokes-Einstein relation. For a liposome of size $R$, the diffusivity is
\begin{equation}
    D = \frac{1}{R}\frac{k_\text{B}T}{6 \pi \eta},
\end{equation}
where $k_{\text{B}}$ is the Boltzmann constant, $T$ is the absolute temperature, and $\eta$ is viscosity. Using the definition $D \equiv \gamma D_0$, where $D_0$ is the on-membrane diffusivity, gives
\begin{equation}
    \gamma = \frac{1}{R}\frac{k_\text{B}T}{6 \pi \eta D_0}.
    \label{eqn:gamma}
\end{equation}

\bibliographystyle{apsrev4-1}

\bibliography{biblio}

\end{document}